\documentclass[a4paper]{jpconf}

\usepackage{citesort}
\usepackage[pdftex]{graphicx}
\begin{document}

\title{ArDM: a ton-scale LAr detector for direct Dark Matter searches}

\author{A. Marchionni$^1$, C. Amsler$^2$, A. Badertscher$^1$, V. Boccone$^2$, A. Bueno$^3$, M. C. Carmona-Benitez$^3$, J. Coleman$^4$, W. Creus$^2$, A. Curioni$^1$, M. Daniel$^5$, E. J. Dawe$^6$, U. Degunda$^1$, A. Gendotti$^1$, L. Epprecht$^1$, S. Horikawa$^1$, L. Kaufmann$^1$, L. Knecht$^1$, M. Laffranchi$^1$, C. Lazzaro$^1$,
P. K. Lightfoot$^6$, D. Lussi$^1$, J. Lozano$^3$, K. Mavrokoridis$^4$, A. Melgarejo$^3$, P. Mijakowski$^7$, G. Natterer$^1$, S. Navas-Concha$^3$, P. Otyugova$^2$, M. de Prado$^5$, P. Przewlocki$^7$, C. Regenfus$^2$, F. Resnati$^1$, M. Robinson$^6$, J. Rochet$^2$, L. Romero$^5$, E. Rondio$^7$, A. Rubbia$^1$, L. Scotto-Lavina$^2$, N. J. C. Spooner$^6$, T. Strauss$^1$, C. Touramanis$^4$, J. Ulbricht$^1$, and T. Viant$^1$}

\address{$^1$ ETH Zurich, Institute for Particle Physics, CH-8093 Z\"{u}rich, Switzerland}
\address{$^2$ Physik-Institut, University of Z\"{u}rich, Winterthurerstrasse190, CH-8057 Z\"{u}rich, Switzerland}
\address{$^3$ University of Granada, Dpto. de Física Teórica y del Cosmos \& C.A.F.P.E, Campus Fuente Nueva, 18071 Granada, Spain}
\address{$^4$ University of Liverpool, Liverpool L69 3BX, United Kingdom}
\address{$^5$ CIEMAT, Div. de Fisica de Particulas, Avda. Complutense, 22, E-28040, Madrid, Spain}
\address{$^6$ University of Sheffield, Department of Physics and Astronomy, Hicks Building, Hounsfield Road, Sheffield, S3 7RH, UK}
\address{$^7$ The Andrzej Soltan Institute for Nuclear Studies, Hoÿza 69, 00-681 Warsaw, Poland}

\ead{alberto.marchionni@cern.ch}

\begin{abstract}
The Argon Dark Matter (ArDM-1t) experiment is a ton-scale liquid argon (LAr) double-phase time projection chamber designed for direct Dark Matter searches. Such a device allows to explore the low energy frontier in LAr with a charge imaging detector. The ionization charge is extracted from the liquid into the gas phase and there amplified by the use of a Large Electron Multiplier in order to reduce the detection threshold. Direct detection of the ionization charge with fine spatial granularity, combined with a measurement of the amplitude and time evolution of the associated primary scintillation light, provide powerful tools for the identification of WIMP interactions against the background due to electrons, photons and possibly neutrons if scattering more than once. A one ton LAr detector is presently installed on surface at CERN to fully test all functionalities and it will be soon moved to an underground location. We will emphasize here the lessons learned from such a device for the design of a large LAr TPC for neutrino oscillation, proton decay  and astrophysical neutrinos searches.
\end{abstract}

\section{Introduction}
\label{sec:intro}
Astronomical observations give strong evidence for the existence of non-luminous and non-baryonic matter, presumably composed of a new type of elementary particles. The leading candidate is a cold thermal relic gas of Weakly Interacting Massive Particles (WIMPs) ~\cite{Steigman:1984ac}, which feel the gravitational interaction, but are otherwise interacting more weakly than standard weak interactions. Direct detection could be achieved by observing the energy deposited when WIMPs elastically scatter from an ordinary target material nucleus, requiring the measurement of recoils of target nuclei with kinetic energy in the range of 5 - 100 keV. 

The Argon Dark Matter (ArDM-1t) experiment ~\cite{Rubbia:2005ge} is using a ton-scale liquid argon (LAr) target for direct WIMP searches, operated as a double-phase time projection chamber with imaging and calorimetric capabilities.

Liquid argon represents a promising target with a favorable form factor and it is only sensitive to spin-independent interactions. It is easily commercially available, since argon constitutes 1\% of air, and it provides adequate self-shielding against the background of low energy photons and electrons. In addition liquid argon is very attractive as detector material, since it has excellent scintillation properties and, given its relatively high electron mobility, is suitable for drifting electrons. Its main properties are summarized in table \ref{TabLAr} and will be discussed more deeply in the following.

\begin{table}
\caption{\label{TabLAr}Main properties of LAr: temperature at 1 atm and corresponding density, attenuation length for 50 keV $\gamma$, electron mobility at the boiling point, average energies needed for the production of an electron-ion pair ~\cite{Miyajima:1974zz} or of one scintillation photon ~\cite{Doke1990617} (measured for 1 MeV electrons), LAr scintillation wavelength, fast and slow scintillation lifetimes.}
\begin{center}
\begin{tabular}{lllllllll}
\br
T & $\rho$ & $\lambda_{att}$ 50 keV $\gamma$ & $\mu_{electron}$ & W$_{ion}$ & W$_\gamma$ & $\lambda_{scint}$ & $\tau_{fast}$ & $\tau_{slow}$\\
at 1 atm (K) & (g/cm$^3$) & (g/cm$^2$) & (cm$^2$/Vs) & (eV) & (eV) & (nm) & (ns) & ($\mu_s$)\\
\mr
 87.2 & 1.396 & 1.4 & 500 & 23.6 & 25 & 128 & 7 & 1.6 \\
\br
\end{tabular}
\end{center}
\end{table}

Ionizing radiation leads to the formation of excited molecular states in either singlet or triplet states, which decay radiatively with fast and slow lifetimes, respectively ~\cite{Hitachi:1983zz}; the lifetimes are largely different, $\tau_{fast}=7$ ns and $\tau_{slow}=1.6$ $\mu$s, allowing to easily discriminate between the two components. The relative abundance of singlet and triplet states depends on the ionization density, and so on the type of ionizing radiation ~\cite{Hitachi:1983zz}. In addition, the phenomenon of recombination, effectively transforming ionization into scintillation ~\cite{PhysRevB.17.2762}, also depends on the ionization density ~\cite{PhysRevA.35.3956,Doke1988291}. Based on these facts, both the time evolution of the argon scintillation light ~\cite{Boulay:2006mb,Lippincott:2008ad} and the relative fraction of scintillation to ionization charge ~\cite{Benetti1993203} are, in LAr, powerful discriminants of nuclear recoil signals against $\gamma$ and electron backgrounds. 

Thanks to the high electron  mobility of LAr, large LAr volumes can be operated as Time Projection Chambers, providing both imaging and energy measurements from the detection of the ionization charge. Relatively high electric fields can be used over extended regions, resulting in very high electric potential values, which are possible in LAr due to its very high dielectric strength. High drift fields could possibly help in the detection of the highly quenched ionization charge from nuclear recoils, as discussed in section \ref{sec:highvoltage}.

A WIMP interaction leading, i.e., to a 30 keV nuclear recoil is expected to produce about 300 VUV photons, together with a few tens of free electrons, both values depending on the electric field strength and still not fully known. This is a very small value for the released charge, which would not be detectable without amplification. A solution is offered by the use of a detector operated in double-phase (liquid-gas) ~\cite{Dolgoshein:1973}, where the ionization charge produced in the liquid is extracted into the gas phase by a sufficiently high electric field at the liquid/gas interface, and there amplified in the pure Ar gas ~\cite{Rubbia:2004tz,Rubbia:2005ge} or observed through proportional scintillation ~\cite{Benetti:2007cd}.

ArDM ~\cite{Rubbia:2005ge} is a one ton argon detector, operated in double-phase, with independent ionization and scintillation readout. The goal is to demonstrate the feasibility of an argon-based ton-scale experiment to efficiently detect WIMP induced nuclear recoils with sufficient background discrimination. Characteristic features of this experiment are the direct charge readout with good spatial resolution, following amplification in the gas phase, and  the possibility to reach high drift fields. The ionization electrons, after drifting in the LAr volume, are extracted into the gas phase and there amplified, directly extrapolating from the more delicate GEM detectors, by thick macroscopic GEMs, also named Large Electron Multipliers (LEMs).

Large liquid argon (LAr) detectors, up to 100 kton scale, are presently being considered for proton decay searches and neutrino astrophysics as well as far detectors for the next generation of long baseline neutrino oscillation experiments, aiming at neutrino mass hierarchy determination and CP violation searches in the leptonic sector ~\cite{Rubbia:2004tz}. The operation of a single LAr module of 100 kton as a charge imaging LAr TPC operated in double phase was first suggested by \cite{Rubbia:2004tz}, in order to compensate the charge loss due to a very long drift with the subsequent amplification of the ionization charge in the pure argon vapor phase. ArDM represents in this respect an important step to assess the feasibility of generation of high drift electric fields and of the amplification and readout of the ionization charge in a LAr TPC operated in double phase.

An overview of the ArDM detector is presented in section \ref{sec:intro}; more detailed descriptions of the cryogenics, scintillation light readout, high voltage generation and ionization charge detection are given in sections \ref{sec:cryo}, \ref{sec:light}, \ref{sec:highvoltage} and \ref{sec:charge}, respectively. The status of the detector, and particularly its novel features, are summarized in section \ref{sec:concl}.

\section{Overview of the detector and prospects}
\label{sec:overview}
\begin{figure}[h]
\begin{minipage}{0.53\linewidth}
\includegraphics[width=0.90\linewidth]{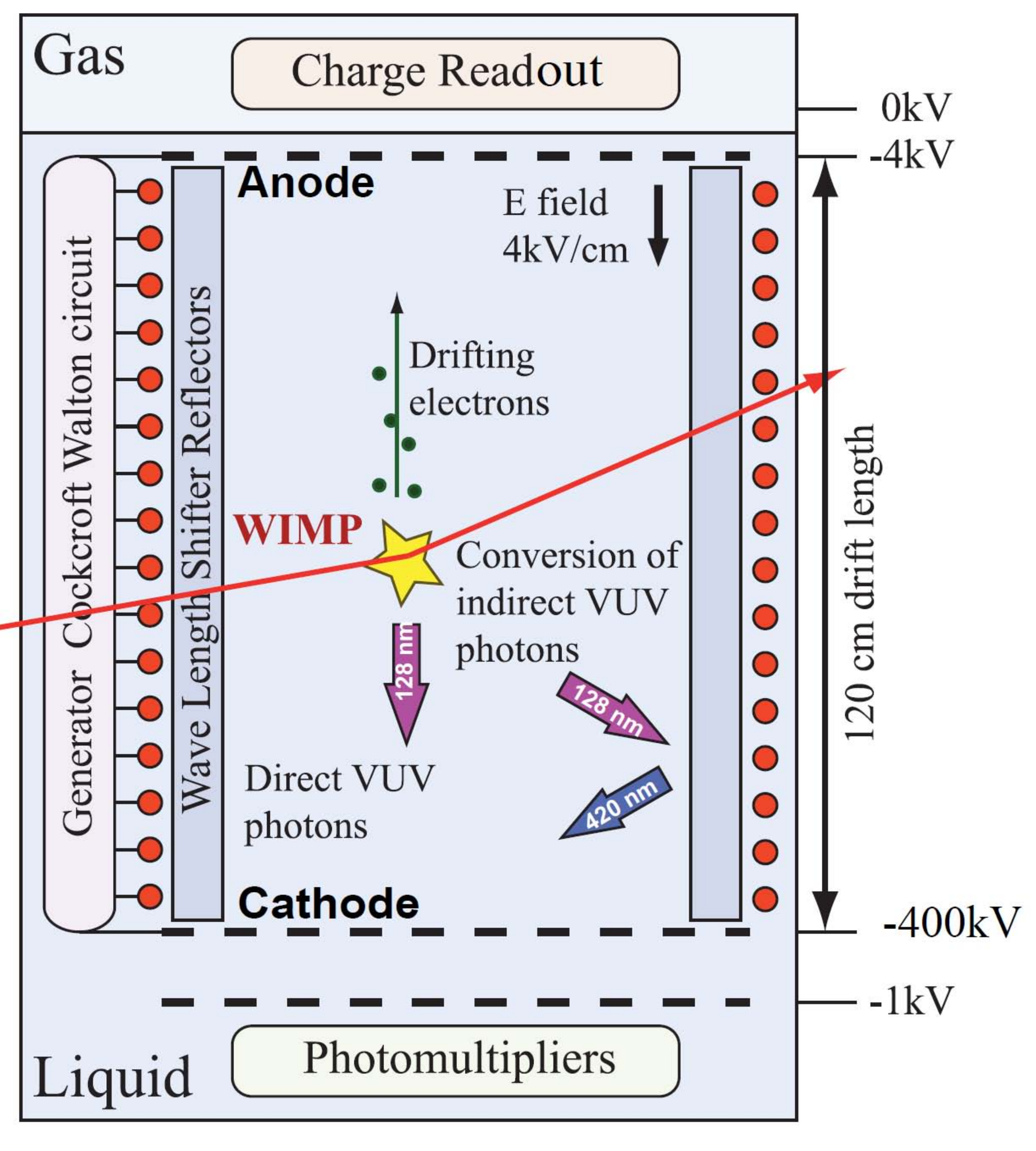}
\caption{\label{ArDMscheme}Conceptual layout of ArDM.}
\end{minipage}\hspace{2pc}%
\begin{minipage}{0.37\linewidth}
\includegraphics[width=0.90\linewidth]{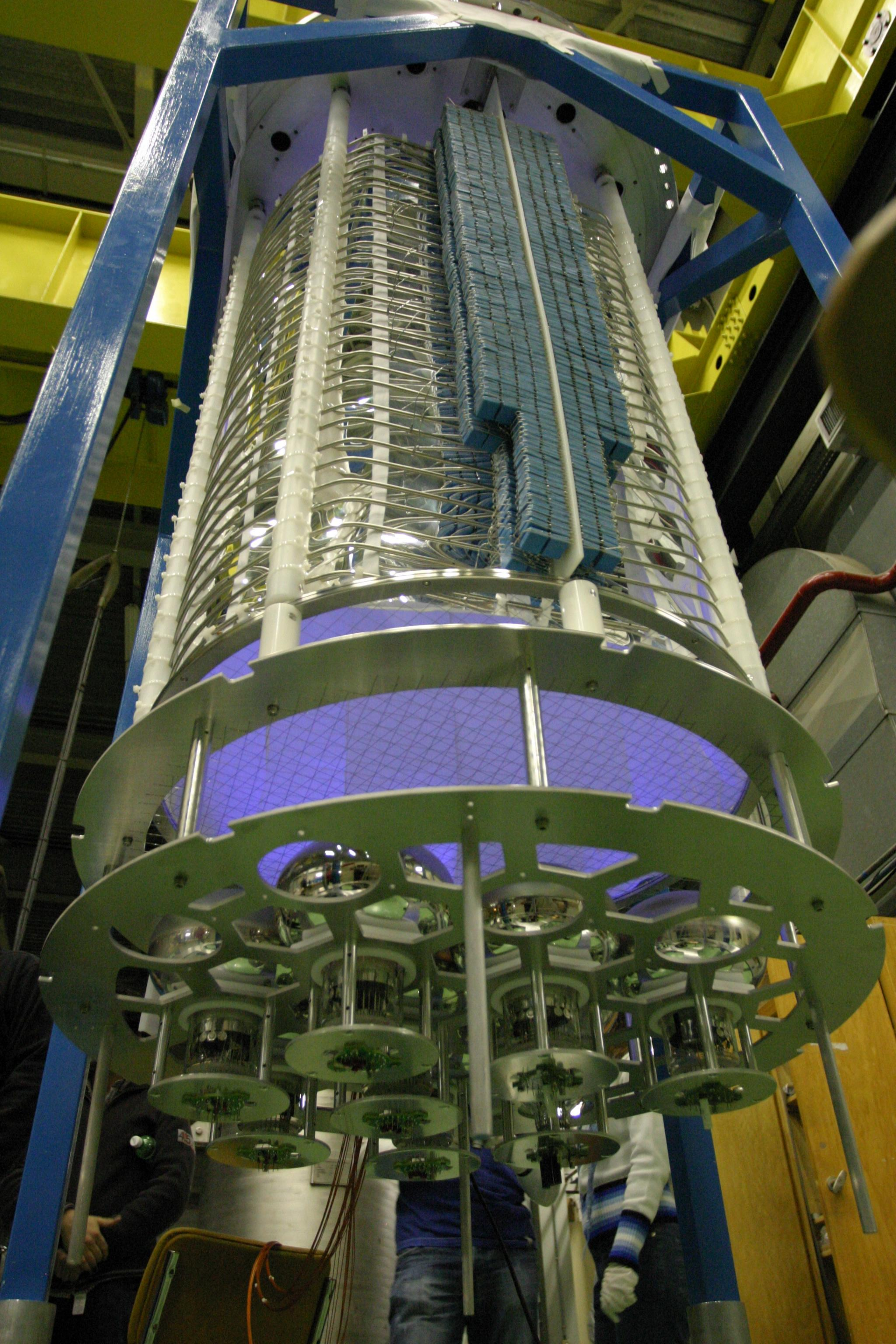}
\caption{\label{ArDMpic}View of the ArDM detector from the bottom}
\end{minipage} 
\end{figure}

Figure \ref{ArDMscheme} shows a conceptual layout of the ArDM apparatus ~\cite{Rubbia:2005ge}. The sensitive liquid argon volume is defined by the drift volume, delimited by a cathode at the bottom and by a set of field shapers on the sides. It has a diameter of 80 cm and a maximum drift length of 120 cm. 

Charged particles lead to production of ionization charge and to excitation of argon atoms, yielding scintillation light emitted isotropicaly along the ionization path at a wavelength of 128 nm. This is converted to blue light (420 nm) by a wavelength shifter (WLS) deposited on foils attached to the field shapers ~\cite{Boccone:2009kk}. The shifted and reflected light is detected by an array of photomultipliers (PMTs) immersed in the liquid at the bottom of the detector, protected from the high electric potential of the cathode by a shielding grid. Direct LAr scintillation light impinging on the PMTs is also detected, since PMTs are coated with WLS.

A 400 kV Cockroft-Walton voltage multiplier ~\cite{Horikawa:2010bv}, immersed in LAr, has some of its stages directly connected to the field shapers, and the last stage connected to the cathode, providing a field, possibly exceeding 3 kV/cm, to drift the electrons to the top surface of the liquid towards the gas phase. Two extraction grids positioned across the liquid-vapour interface generate a $\approx$ 3 kV/cm field to extract the electrons to the gas phase, where they are driven inside the holes of a Large Electron Multiplier which provides charge multiplication ~\cite{Badertscher:2008rf}. The amplified charge is finally collected by a two-dimensional projective anode with fine spatial resolution ~\cite{Badertscher:2010zg}.

Figure \ref{ArDMpic} shows a photograph of the detector, displaying, from the bottom to the top, the cryogenic photomultipliers with the shielding grid on top, the cathode, the set of field shapers and, hanging from the top, the Cockroft-Walton generator. In addition it shows the wavelength shifting foils, while illuminated with a UV lamp.

The $^{39}$Ar, a beta-emitter with a lifetime of 269 years and a Q-value of 565 keV, naturally present in the atmospheric argon ~\cite{Loosli198351,Benetti:2006az} and causing a background rate of about 1 kHz in ArDM, will eventually be the limiting factor in the sensitivity of the experiment. It may be possible to replace the atmospheric argon contaminated with $^{39}$Ar with argon extracted from underground wells, shown to be depleted from $^{39}$Ar by more than an order of magnitude ~\cite{AcostaKane200846}. 

The ArDM detector is presently installed on surface at CERN to fully test all functionalities and it will be later moved to an underground location. ArDM could possibly investigate WIMP cross sections down to $10^{-45}$ cm$^2$, depending on background rejection. Within a "canonical" WIMP halo mode, a WIMP with a mass of 100 GeV and a cross-section of $10^{-44}$ cm$^2$ would yield about 1 recoil event per day per ton of argon above a recoil energy threshold of 30 keVr.

\section{Cryogenics and safety}
\label{sec:cryo}
\begin{figure}[h]
\begin{minipage}{0.50\linewidth}
\includegraphics[width=0.90\linewidth]{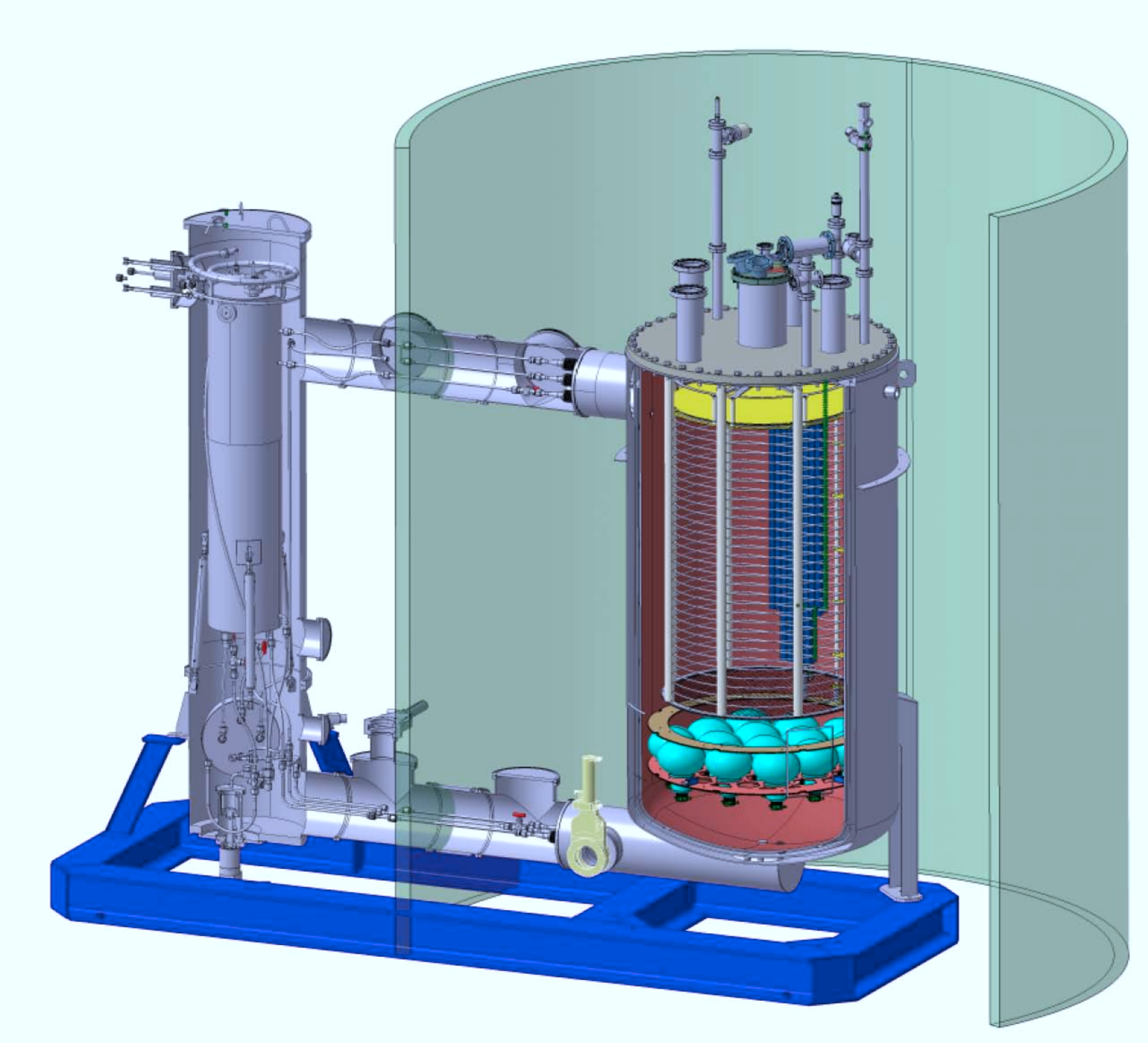}
\caption{\label{detcryo}Schematics of the detector vessel (on the right), surrounded by the radiation shield, and of the purification column (on the left).}
\end{minipage}\hspace{2pc}%
\begin{minipage}{0.50\linewidth}
\includegraphics[width=0.90\linewidth]{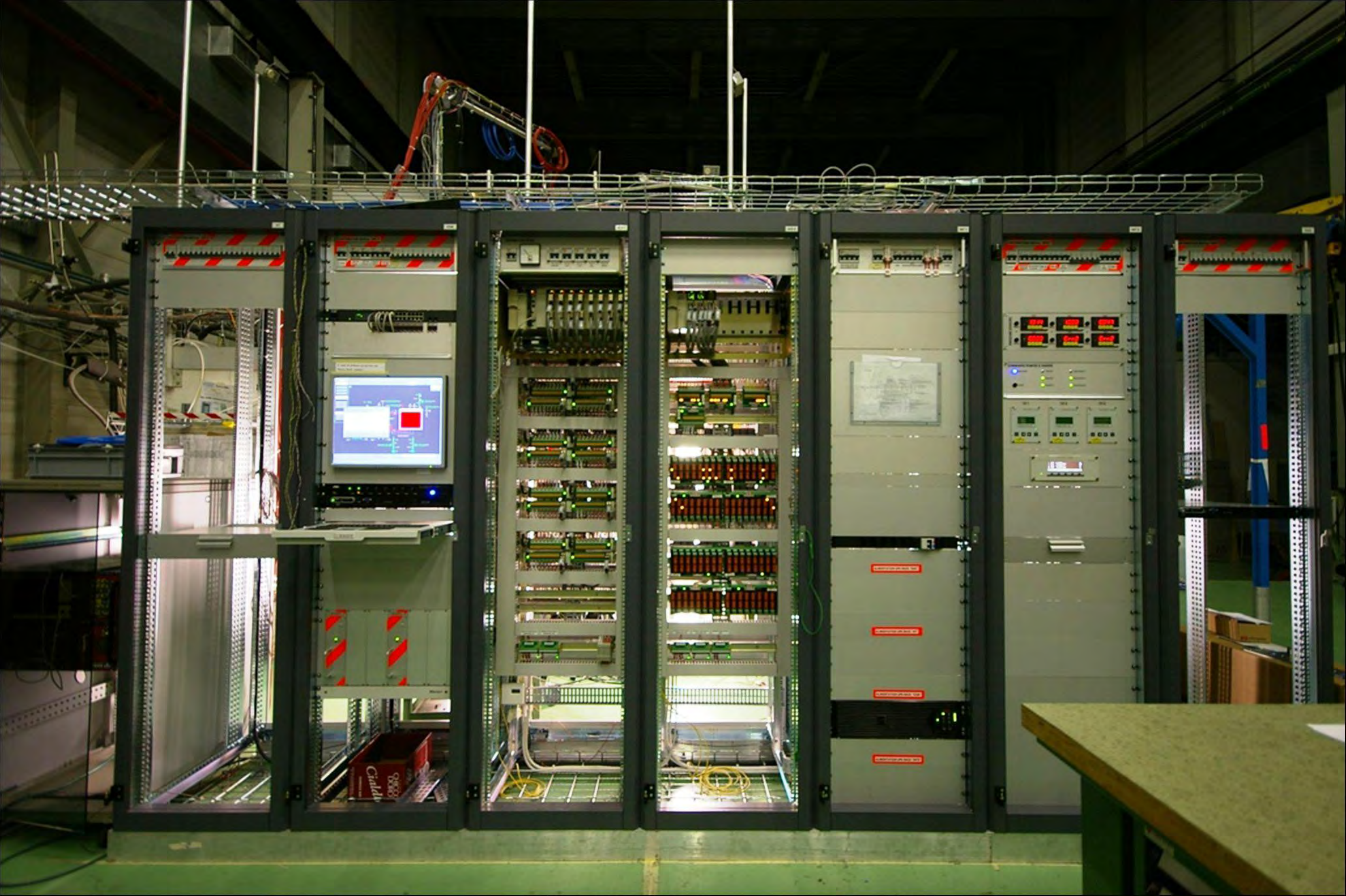}
\caption{\label{PLC}View of the electronic equipment racks of ArDM, including the PLC system and vacuum and cryogenic controls.}
\end{minipage} 
\end{figure}

The detector is housed in a vacuum tight stainless steel dewar vessel, with an inner diameter of 100 cm and a height of about 2 m, containing approximately 1.4 m$^3$ of LAr, along with a layer of saturated argon vapor at the top. Before filling with LAr, the detector vessel is evacuated to a high vacuum, below 10$^{-6}$ mbar, to minimize the contamination of the liquid from remaining impurities and outgassing.  The purity of the liquid argon is maintained by a custom-made purification cartridge, based on CuO powder, separated from the main dewar to allow the insertion of a radiation shield against neutrons (see figure \ref{detcryo}). A cryogenic pump allows recirculation of the LAr in the detector vessel through the purification cartridge. 

The detector vessel and the purification cartridge are fully surrounded by a jacket of LAr, named argon bath, which is used for the initial cooling of the detector and for maintaining stable cryogenic conditions. The surrounding LAr jacket is enclosed by an outer shell of stainless steel, where in between it is maintained a vacuum better than $10^{-4}$ mbar for thermal insulation.  Presently the LAr bath is open to the atmosphere so that the temperature is maintained at 87 K, corresponding to a saturated argon vapor pressure of 1 bar. This implies that the pressure of the saturated gas on top of the liquid in the detector vessel is also kept at atmospheric pressure.

For the future we want also to operate the LAr bath as a closed system and keep the temperature stable by the use of cryocoolers. This allows more flexibility in the choice of the temperature, possibly cooling below a saturated argon vapor pressure of 1 bar, with no dependance on variations of the atmospheric pressure. It also removes the need for a continual refilling of the LAr bath, impracticable for a continuous operation in an underground environment. The needed cooling power, between 300 and 400 W at LAr temperature, has been estimated by measuring the LAr consumption when operating in the open LAr bath configuration. It is dominated by the heat input through the top flange, which is not equipped with an insulating vacuum. We are in the process of installing two $\sim 250$~W crycoolers, with enough flexibility to allow operation in a subcooled state. 

As the detector contains such a large amount of cryogenic liquid in a
closed volume, it is necessary to protect it with the aid of pressure relief
devices. Since pressure relief valves do not meet the tight leak rate requirements needed to maintain the purity of the LAr in the detector vessel, the system has been equipped with two burst discs, with a burst pressure of ~0.7 barg. Pressures and temperatures of the system, together with the vacuum levels for thermal insulation, are continuously monitored through a Programmable Logic Controller (PLC) system. The PLC controls all the vacuum and cryogenic processes: operation of the vacuum pumps, recirculation of the LAr in the dector vessel, refilling of the LAr bath and in the future the operation of the cryocoolers. A view of the ArDM electronic equipment racks, also housing the PLC system and all vacuum and cryogenic controls, is shown in figure \ref{PLC}. 

\section{Light readout with cryogenic PMTs and wavelength shifting techniques}
\label{sec:light}
The emission spectrum of LAr scintillation is peaked at 128 nm, which would require the use of VUV sensitive PMTs (e.g. MgF$_2$ windowed), not commercially available with large area photocathodes. An alternative and still efficient solution is the use of reflectors coated with a wavelength shifter (WLS) coupled with standard bialkali photomultiplier tubes. In ArDM ~\cite{Boccone:2009kk} the reflectors consist of Tetratex (TTX) foils, 254 $\mu$m thick, coated by evaporation with 1 mg/cm$^2$ of TetraPhenyl Butadiene (TPB), which presents a fluorescence decay time of 1.68 ns and an emission spectrum between 400 and 480 nm. A reflection coefficient close to 97\% has been measured for evaporated TTX samples. 

\begin{figure}[h]
\begin{minipage}{0.567\linewidth}
\includegraphics[width=0.90\linewidth]{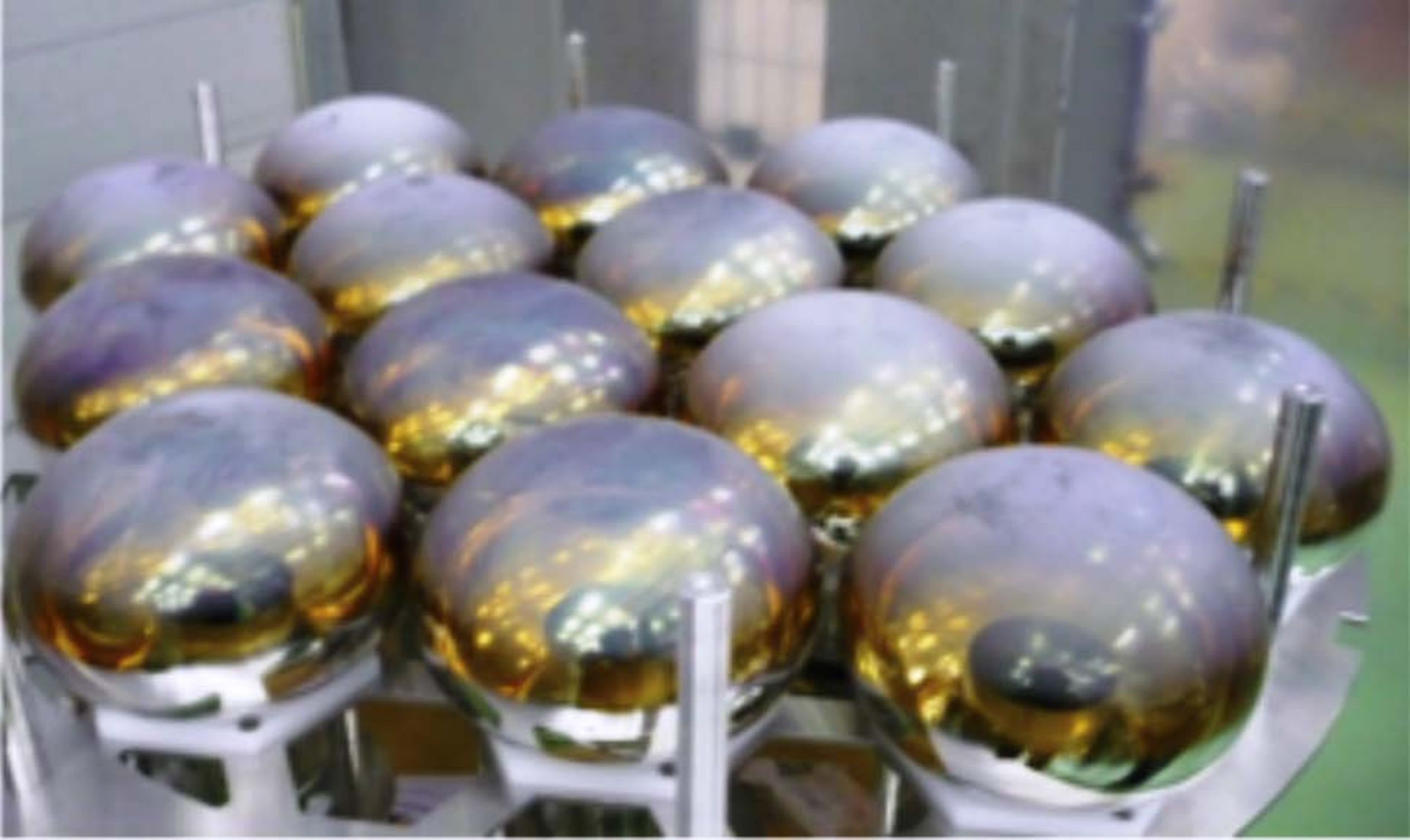}
\caption{\label{ArDMPMT}A photograph of the assembly of the 14 8" PMTs on a
 stainless steel holder plate.}
\end{minipage}\hspace{2pc}%
\begin{minipage}{0.333\linewidth}
\includegraphics[width=0.90\linewidth]{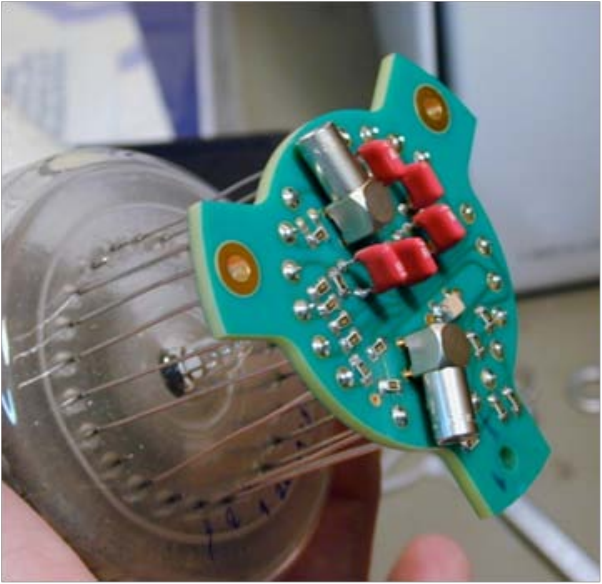}
\caption{\label{Cryobase}View of a cryogenic voltage divider.}
\end{minipage} 
\end{figure}

The shifted and reflected light is collected by 14 PMTs (8" Hamamatsu R5912-02MOD with bialkali photocathodes and Pt-underlay, see figure \ref{ArDMPMT}) located at the bottom of the cryostat and completely immersed in LAr. The PMTs were also coated by evaporation with a thin layer (0.05 mg/cm$^2$) of TPB for direct scintillation light detection. Each PMT is soldered with its leads onto a 3 mm thick printed circuit board (see figure \ref{Cryobase}), providing the voltage divider for cathode and dynodes, using only passive electronic components. A 5 mm thick stainless steel plate with hexagonal holes serves as a holder for the PMTs. The PMTs are firmly held in place by simply attaching the printed boards to the support plate.

\begin{figure}[h]
\includegraphics[width=0.90\linewidth]{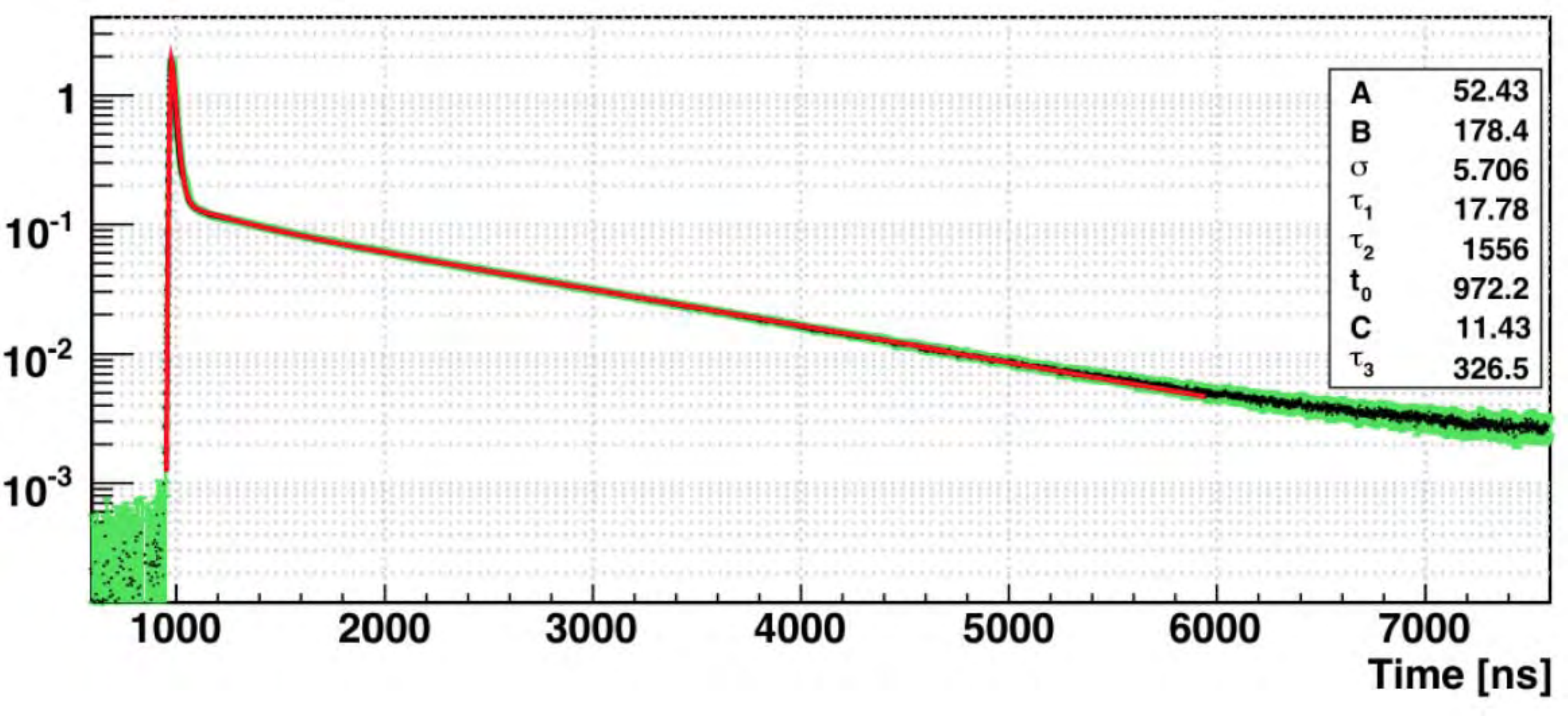}
\caption{\label{PMTsignal}Typical LAr scintillation signal induced by minimum ionizing particles.}
\end{figure}

The PMTs were calibrated by using short light pulses from a 400 nm blue LED, placed in the vapor region above the liquid argon surface. The PMT gains were set to $2.5 \times 10^7$ by individually adjusting the bias voltages. A typical pulse shape for scintillation in LAr induced by electrons, gammas and cosmic muons, selected by requiring a fast/slow component $\le 0.4$, is shown in figure \ref{PMTsignal}. A fit to the signal ~\cite{Boccone:2010} shows the necessity of an intermediate component, providing lifetimes of $\tau_1=17.8 \pm 1.0$ ns, $\tau_2=1556 \pm 10$ ns, $\tau_3=326 \pm 10$ ns, and corresponding contributions of 21, 74 and 5\%, for the fast, slow and intermediate components, respectively. No attempt was made to extract the fast component LAr lifetime by correcting for the time response of the PMTs and of the data acquisition system. 

A preliminary set of only seven PMTs were tested in LAr with ArDM on surface at CERN in 2009. By using external radioactive sources it was measured a light yield at zero electric field between 0.3-0.5 phe/keVee depending on the position
within the detector volume ~\cite{Amsler:2010yp}. A new recent test performed in 2010 with a completed detector with 14 PMTs is still being analyzed.
 
During these tests of ArDM we found very convenient to monitor the LAr purity by measuring the $\tau_2$ lifetime. The $\tau_2$ lifetime is quite sensitive to traces of impurities, down to 0.1 ppm, allowing to easily monitor a deterioration of the purity. In particular we use this method to quickly detect possibly contaminated LAr batches at the delivery time. This is particularly important when filling large detectors, which require the delivery of many LAr batches, and the addition of only one bad batch could seriously deteriorate the LAr purity of the whole detector. In ArDM we measured the $\tau_2$ lifetime over about 600 h with no significant deterioration, allowing to set a limit of $\le 0.1$ ppm for the N$_2$ and O$_2$ concentrations ~\cite{Boccone:2010,Acciarri2009169}.

\section{Development of High Voltage systems}
\label{sec:highvoltage}
The development of high voltage systems is crucial for LAr TPCs, which need a stable and uniform drift field over extended LAr volumes. In case of ArDM the maximum drift length is 120 cm, but drift lengths up to 20 m have been envisaged for future large LAr TPCs ~\cite{Rubbia:2004tz}. There are clear advantages in going to higher drift electric fields:
\begin{itemize}
\item shorter charge collection times, in order to reduce the attenuation of the drift electrons due to their attachment to electronegative impurities present in LAr, e.g. oxygen. In fact the ratio of collected to released ionization charge is well described by $Q/Q_0 = e^{-t/\tau}$, where t is the drift time and $\tau$ the electron lifetime, inversely proportional to the concentration of impurities according to $\tau [\mu s] = 300/\rho_{O_2} [ppb]$. In LAr the drift velocity of electrons increases less than linearly for electric fields above 100 V/cm, still there is considerable gain at higher fields. It reaches 1.55 mm/$\mu$s at 0.5 kV/cm for a temperature T=89 K, and increases by 30\% by doubling the electric field intensity from 0.5 to 1 kV/cm, and again by 30\% from 1 to 2 kV/cm ~\cite{Walkowiak:2000wf}.
\item Attachment cross sections of electrons to O$_2$ impurities in LAr have been measured in ~\cite{Bakale1976} to decrease as a function of the electric field . Anyhow in ~\cite{Amoruso:2004ti} it was found only a weak dependance of the attachment cross-section on electric fields in the range 0.05-1 kV/cm.
\item Recombination of electrons and positive ions from the initial ionization charge decreases with increasing electric fields, particularly for heavily ionizing particles. Measurements in LAr have been reported by ~\cite{Gruhn1978,Scalettar1982,Shibamura1987437,Thomas1987,Aprile1987519,Cennini:1994ha,Amoruso:2004dy}. In reference ~\cite{Amoruso:2004dy} recombination factors for relativistic particles with unit charge are fitted with a Birks law dependance as a function of the ratio of the particle stopping power dE/dx to the applied drift electric field. This is purely a phenomenological expression and, as cautioned in ~\cite{Amoruso:2004dy}, it is not valid in the high recombination region, that is in either low electric fields or high ionization density. The dependance of recombination on the electric field is more pronounced in LAr than LXe ~\cite{Thomas1987}. In LXe it was also found an almost complete field independence of the ionization yield for nuclear recoils ~\cite{Aprile2006} , a striking result not completely understood. The possibility to go to high drift electric fields in ArDM, up to 3 kV/cm, will be important to experimentally assess recombination effects as a function of electric field for nuclear recoils.
\end{itemize}

ArDM has a cylindrical drift volume of 80 cm diameter and 120 cm height, delimited by a cathode grid at the bottom and by 30 ring-shaped electrodes, called field shapers, spaced by 40 mm. The solution adopted in ArDM for the high voltage generation is to have a 210 stages Cockcroft-Walton voltage multiplier directly immersed in LAr ~\cite{Horikawa:2010bv} (see figure \ref{ArDMpic}), driven by a 50 Hz AC charging voltage with a maximum V$_{pp}=2.5$ kV. Each capacitor in the Cockroft-Walton scheme is made out of four (two) 82 nF metallized film polypropylene capacitors in parallel for stages 1 to 170 (171 to 210) and each diode is realized by three high voltage avalanche diodes in series. The redundancy built in the system prevents from a critical loss of functionality of the circuit. The voltage multiplier can be operated at a low frequency charging voltage, well outside of the relevant bandwidth of the charge and light readout electronics, since a negligible current is drawn by the voltage multiplier: in fact the DC voltages generated by the circuit are directly connected to the cathode and to the field shapers. Given the large number of Cockroft-Walton stages, it is possible to choose appropriately the connections to the field shapers in order to achieve a good linearity in the voltage distribution. 

Voltage multipliers with a large number of stages are affected by the shunt capacitance due to the diode capacitance and the stray capacitance between two adjacent stages, causing the output voltage to increase less than linearly with the stage number, even with no current load. In case of ArDM this non-linearity has been measured, resulting in a voltage on the last stage of 72\% of the maximum theoretical attainable voltage. This results in a cathode voltage of 378 kV, when driving the circuit at V$_{pp}=2.5$ kV, and a drift electric field exceeding 3 kV/cm. This solution allows to reach interestingly high values for the drift electric field, in order to reduce the charge attenuation due to the presence of electronegative impurities in LAr and to limit the phenomenon of recombination. Externally generated high voltages would need the use of cryogenic feedthroughs, presently tested only up to 150 kV ~\cite{Amerio:2004ze}. 

In a preliminary test in ArDM, the Cockroft-Walton voltage multiplier has been successfully operated up to 70 kV. 

\section{Charge collection and amplification}
\label{sec:charge}
The detection of a few tens of ionization electrons is one of the most challenging aspects of the ArDM detector. The sensitive volume of the detector is defined by the drift volume, already described in section ~\ref{sec:highvoltage}. The ionization electrons are drifted towards the liquid surface on top, possibly with a high drift electric field value to reduce attenuation due to electronegative impurities in LAr, are extracted into the gas phase by means of two extraction grids precisely positioned across the liquid, and then amplified in the pure Ar gas by thick macroscopic GEMs called Large Electron Multipliers. 

\begin{figure}[h]
\begin{minipage}{0.36\linewidth}
\includegraphics[width=0.90\linewidth]{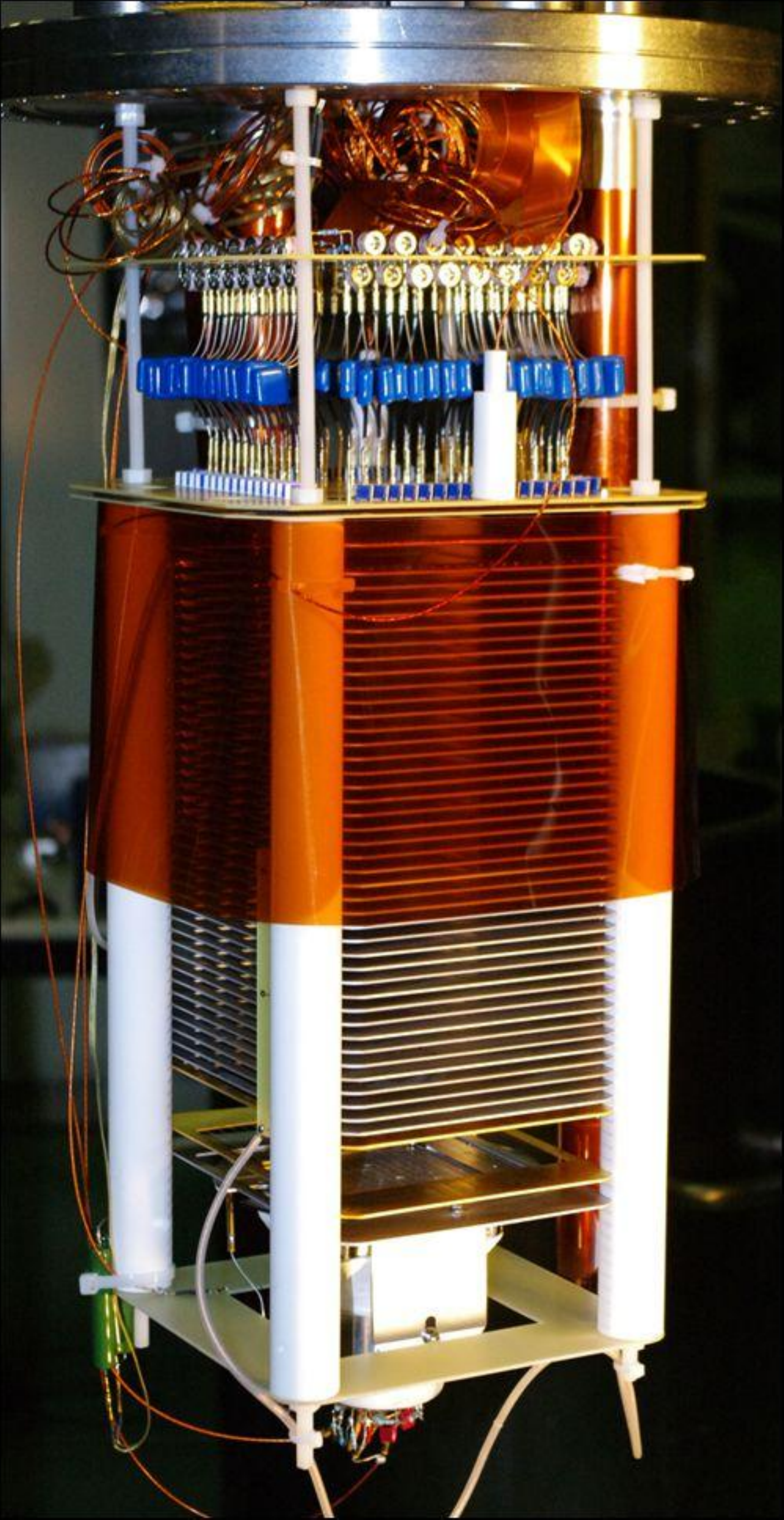}
\caption{\label{LEMsetup}Picture of the LAr LEM-TPC test setup with an active area of $10 \times 10$~cm$^2$ and a drift length of 21 cm.}
\end{minipage}\hspace{2pc}%
\begin{minipage}{0.54\linewidth}
\includegraphics[width=0.90\linewidth]{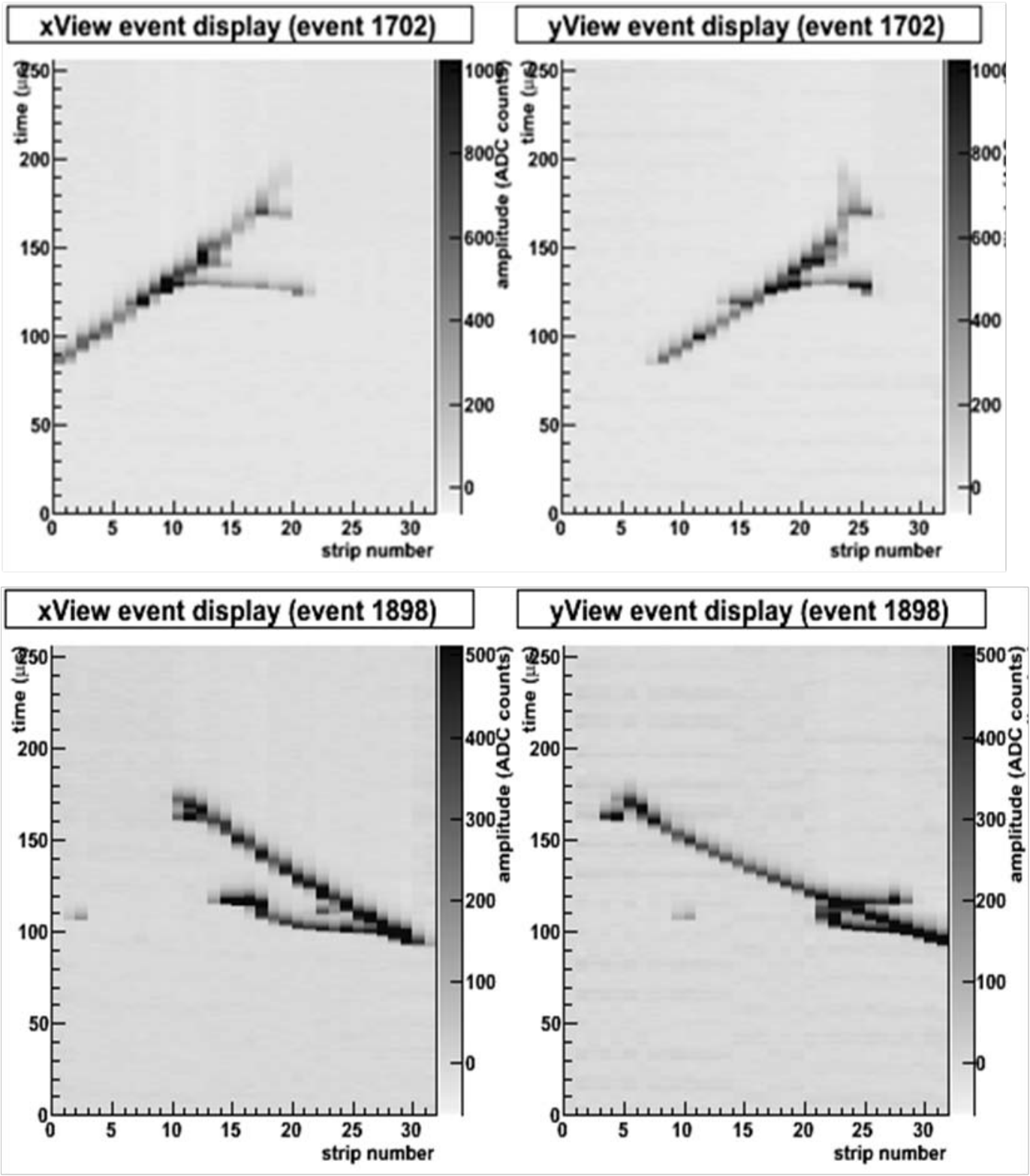}
\caption{\label{LEMevent}LEM-TPC displays of two muon events (top and bottom pictures) crossing the LAr sensitive volume with visible delta rays: left) drift time vs x-channel number; right) drift time vs y-channel number.}
\end{minipage} 
\end{figure}

In order to get charge amplification in pure argon gas, without any quencher, it is necessary to confine the electron avalanche, which would be otherwise spread by the argon scintillation photons produced in the high electric field regions by proportional scintillation. The geometry of the GEM is particularly suited to be used in pure Ar gas, because it generates the high electric fields, necessary for charge amplification, only inside the holes, effectively confining the electron avalanche. GEMs have been successfully tested in pure Ar gas at normal pressure and temperature, obtaining gains of the order of 1000 ~\cite{Bressan:1998et}. They have also been tested in pure Ar in double phase conditions using triple-stage GEMs, reaching gains of the order of 5000 ~\cite{Bondar:2005wx}, in order to compensate for the essentially inverse exponential dependance of the first Townsend coefficient with gas density ~\cite{Aoyama1985125}. Large Electron Multipliers (or Thick GEMs) ~\cite{Otyugova:2008,Bondar:2008yw} with mm size holes in a mm thick printed circuit board, are a natural evolution of GEMs on a more rigid structure better suited for a cryogenic environment.

In order to develop a LEM for the ArDM detector, we started a series of tests on LEMs of different thickness and manufacturing procedure and with different schemes for the charge readout. We constructed a 3 l active volume double-phase LAr TPC with a LEM readout system, as shown schematically in figure ~\ref{LEMsetup}, which we call LAr LEM-TPC. Details on the construction and operation of the device are presented in ~\cite{Badertscher:2008rf,Badertscher:2009av,Badertscher:2010fi,Badertscher:2010zg}.
In order to get a full 3-dimensional reconstruction of an ionizing event, it is necessary, in addition to the drift time information, to induce signals on at least two independent spatial views, typically realized by two sets of orthogonal strips in case of a bi-dimensional readout. We found important to decouple the mechanism of charge amplification in the LEM from the charge readout, so we chose to drift the amplified charge to a two-dimensional projective readout anode, providing two independent views with 3 mm spatial resolution ~\cite{Badertscher:2010zg}. With a single 1 mm thick LEM, coupled to a two-dimensional projective anode, we achieved an effective gain of 27 on the released ionization charge, without any degradation in charge collection resolution. Examples of recorded events are shown in figure ~\ref{LEMevent}, displaying muon events crossing the sensitive volume, accompanied by delta rays. A signal to noise ratio larger than 200 has been obtained for minimum ionizing particles, thanks to charge amplification, resulting in images of very high quality.

For the charge detection of WIMP initiated nuclear recoils in ArDM, corresponding to an ionization charge down to a few tens of electrons, charge gains of about 500 are necessary, since a signal to noise ratio of 10 for 1 fC input charge is typical for our charge preamplifiers. Such gains should be achievable in setups with two 1 mm thick LEMs operated in cascade, as it will be soon verified in an experimental test.

\begin{figure}[h]
\begin{minipage}{0.45\linewidth}
\includegraphics[width=0.90\linewidth]{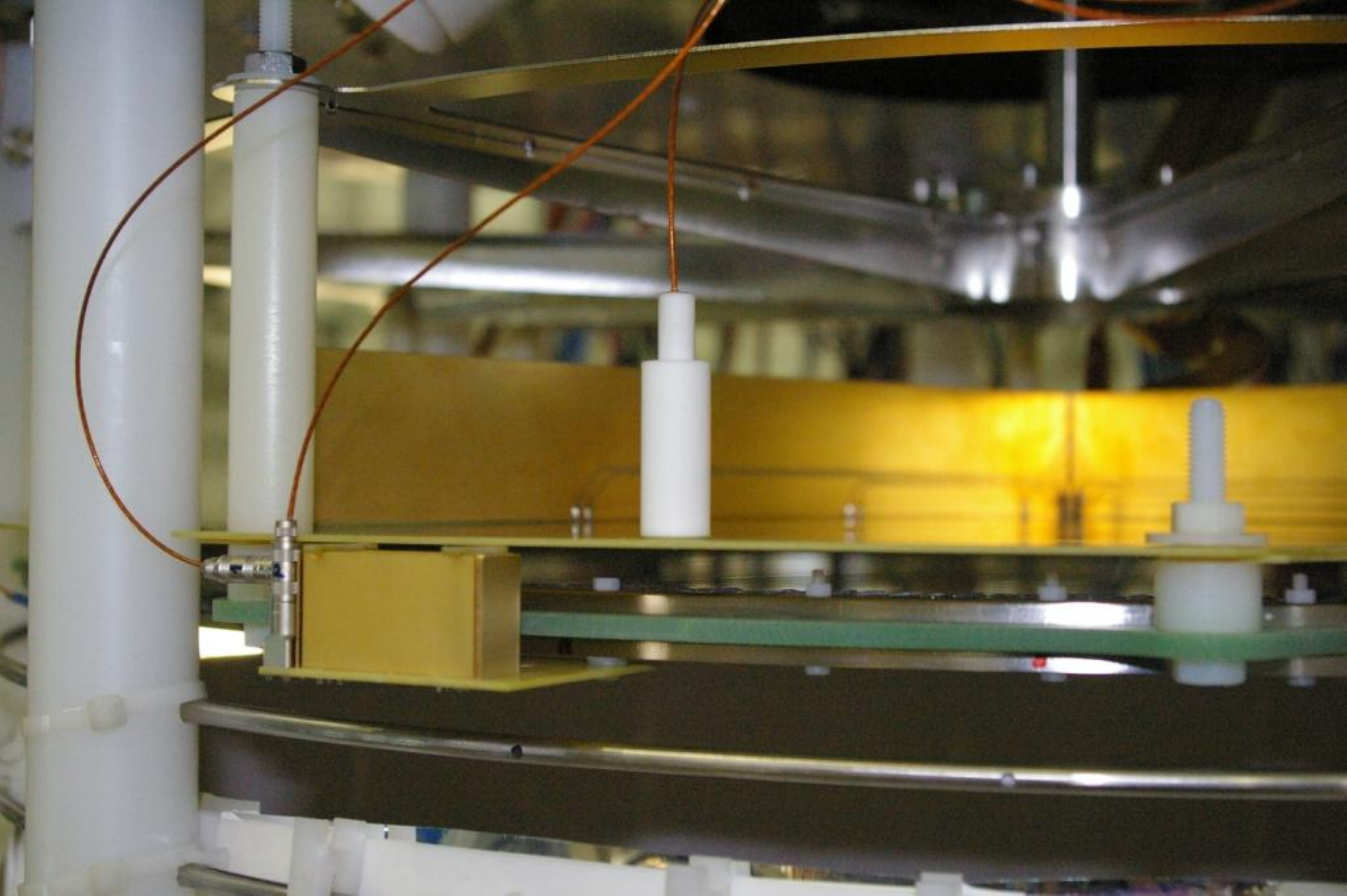}
\caption{\label{ChargeReadoutAssembly}View of the charge readout assembly.}
\end{minipage}\hspace{2pc}%
\begin{minipage}{0.45\linewidth}
\includegraphics[width=0.90\linewidth]{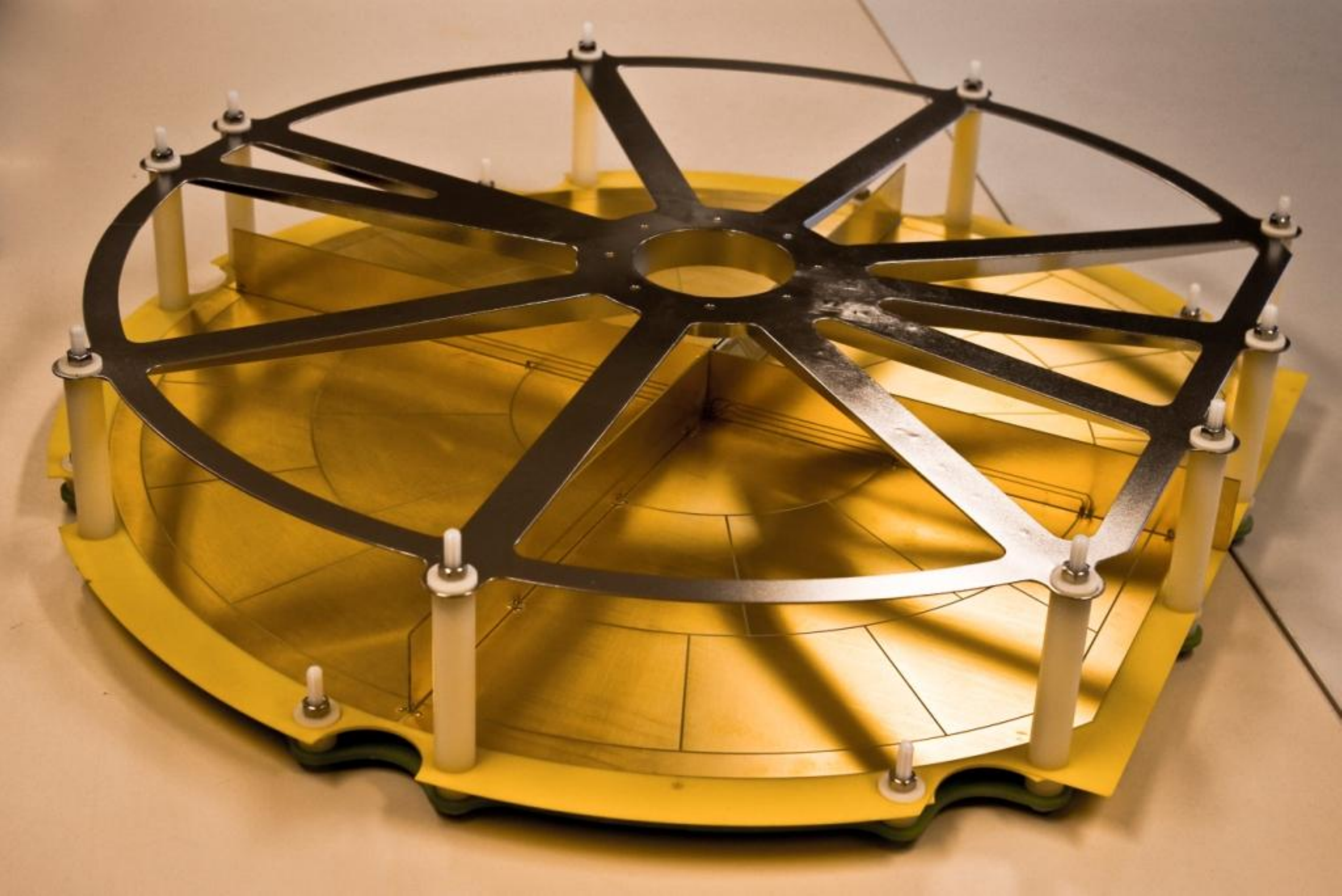}
\caption{\label{ChargeReadoutPlane}Picture of the temporary charge readout electrode, subdivided into 32 pads.}
\end{minipage} 
\end{figure}

Figure ~\ref{ChargeReadoutAssembly} is a view of the present charge readout assembly in ArDM, showing, from bottom to top, the upper field-shaper, the two extraction grids precisely kept at 10 mm distance by a fiberglass ring spacer, and the charge readout electrode, positioned 10 mm above the upper grid. Also visible is one of three capacitive level meters, directly mounted on the extraction grids, to monitor the liquid level between the two grids. The whole assembly is hanging from a mechanical support capable of vertical and tilting movements, adjustable from outside the dewar, which allows a precise positioning and leveling of the extraction grids across the liquid-vapour interface.

In a first test of drift and extraction of the ionization charge, the Cockroft-Walton voltage multiplier has been operated at 70 kV, corresponding to an electric drift field of $\sim 0.5$ kV/cm. For this test a temporary electrode with 32 pads readout, with no charge amplification, has been used for the collection of the extracted charge, as shown in figure ~\ref{ChargeReadoutPlane}. The same pad structure is present on the lower and upper faces of the charge readout plane. Charge signals are readout from the upper pads, capacitatively coupled to the lower pads, which are kept at high voltage. By triggering on muons crossing the LAr sensitive volume by means of external scintillator plates, both scintillation light and ionization charge have been detected. In addition proportional scintillation light from the extraction of the ionization charge into the gas phase, due to the high electric field region between the two extraction grids, has been observed.

\section{Conclusions}
\label{sec:concl}
ArDM, a one ton liquid argon double-phase TPC, has been assembled at CERN and it will soon be moved to an underground location to be operated as Dark Matter detector. It relies on the detection of both primary scintillation light and ionization charge with good spatial resolution for the identification of nuclear recoils initiated by WIMP interactions. 

Novel features of ArDM are the possible use of high electric fields, generated by a Cockroft-Walton voltage multiplier directly immersed in LAr, and the direct readout of the ionization charge, extracted from the liquid to the gas phase, and there amplified by the use of Large Electron Multipliers. The light detection system, recently completed with 14 PMTs, has been experimentally shown to be able to detect low energy deposits as those expected from nuclear recoils. In a first test, the Cockroft-Walton multiplier has been operated stably at 70 kV, resulting in an electric drift field of $\sim 0.5$ kV/cm. The drifted charge has been extracted into the gas phase by means of two extraction grids, precisely positioned across the liquid level. Proportional scintillation light from the extraction of the ionization charge into the gas phase, due to the high electric field region between the two extraction grids, has been observed. A temporary electrode with pads readout, with no charge amplification, has been used for the collection of the extracted charge. 

Successful charge extraction from the liquid argon phase and multiplication in the gas phase have been routinely achieved in a smaller test setup by using Large Electron Multipliers. An effective gain of 27 on the released ionization charge has been reached with a single 1 mm thick LEM of $10\times 1$0 cm$^2$. The charge is finally collected by a two-dimensional projective anode with 3 mm spatial resolution. Gains of about 500, as necessary for the charge detection of WIMP initiated nuclear recoils, should be possible in a setup with two 1 mm thick LEMs operated in cascade, and it will be soon verified in an experimental test. We do not foresee any fundamental problem in the construction of a $\sim 1$ m$^2$ LEM, as necessary for ArDM, and we are expecting soon the delivery of a $\sim 0.3$ m$^2$ system for a different application.

ArDM will explore the low energy frontier in LAr with a charge imaging device, but at the same time it represents an important step to test some of the features of the design of very large LAr TPCs ~\cite{Marchionni:2009tj}.
\section*{References}
\bibliography{ArDMbib}
\bibliographystyle{iopart-num}

\end{document}